\input amstex.tex
\documentstyle{amsppt}
\newcount\refcount
\advance\refcount 1
\def\newref#1{\xdef#1{\the\refcount}\advance\refcount 1}
\newref\knilllaflamme
\newref\macwilliamssloane
\newref\shadowenum
\newref\shorlaflamme
\def\supp{\operatorname{supp}}
\def\Hom{\operatorname{Hom}}
\def\wt{\operatorname{wt}}
\def\Tr{\operatorname{Tr}}
\topmatter
\title Quantum weight enumerators \endtitle
\author Eric Rains\endauthor
\affil AT\&T Research \endaffil
\address AT\&T Research, Room 2D-147, 600 Mountain Ave.
         Murray Hill, NJ 07974, USA \endaddress
\email rains\@research.att.com \endemail
\date November 4, 1996 \enddate
\abstract
In a recent paper ([\shorlaflamme]), Shor and Laflamme
define two ``weight enumerators'' for quantum error correcting codes,
connected by a MacWilliams transform, and use them to give a
linear-programming bound for quantum codes.  We introduce two new
enumerators which, while much less powerful at producing bounds,
are useful tools nonetheless.  The new enumerators are connected by a
much simpler duality transform, clarifying the duality between Shor and
Laflamme's enumerators.  We also use the new enumerators to give a
simpler condition for a quantum code to have specified minimum distance,
and to extend the enumerator theory to codes with block-size greater than 2.
\endabstract
\endtopmatter

\head Introduction\endhead

One of the basic problems in the theory of quantum error correcting codes
(henceforth abbreviated QECCs) is that of giving good upper bounds on the
minimum distance of a QECC.  The strongest technique to date for this
problem is the linear programming bound introduced by Shor and Laflamme
([\shorlaflamme]).  Their bound involves the definition of two ``weight
enumerators'' for a QECC; the two enumerators satisfy certain inequalities
(e.g., nonnegative coefficients), and are related by MacWilliams
identities.  This allows linear programming to be applied, just as for
classical error correcting codes ([\macwilliamssloane]).

We introduce two new enumerators, called unitary enumerators, with simple
definitions, manifestly invariant under equivalences of quantum codes.
This leads to simpler conditions for codes to have specified minimum
distance.  Moreover, the duality between the unitary enumerators is much
simpler than the duality between the Shor-Laflamme enumerators, hopefully
helping to clarify the nature of that duality.

The final benefit of the unitary enumerators is that they generalize easily
to block quantum codes (codes in which the basic unit has more than two
states), allowing all of the enumerator machinery to be applied there as
well.

Section 1 reviews the Shor-Laflamme enumerators and proves some basic
results.  Section 2 defines the unitary enumerators, shows how they are
related to the Shor-Laflamme enumerators, and proves a number of results,
including duality and minimum distance criteria.  Section 3 extends
everything to block quantum codes, first extending the unitary enumerators,
then the Shor-Laflamme enumerators.  Section 4 states a conjecture
concerning the extension of a fifth enumerator ([\shadowenum]) to
block codes.  Finally, section 5 uses the new minimum distance criteria
to analyze some ways to construct new quantum codes from old quantum codes,
including, in particular, concatenation of codes.

A quick note on terminology: We will be using the terms ``pure'' and
``impure'' in place of the somewhat cumbersome terms ``nondegenerate'' and
``degenerate''; that is, a pure code is one in which all low weight
errors act nontrivially on the codewords.

\head 1. The Shor-Laflamme enumerators ($A$ and $B$) \endhead

Recall that a quantum code $\Cal{C}$ is a $K$-dimensional subspace of a
$2^n$-dimensional Hilbert space $V$; $\Cal{C}$ has minimum distance at
least $d$ if and only if
$$
\langle v|U_{d-1}|v\rangle=\langle w|U_{d-1}|w\rangle,
$$
for $v$ and $w$ ranging over all unit vectors in $\Cal{C}$
([\knilllaflamme]), and for $U_{d-1}$ ranging over all $d-1$ qubit errors.
We will use the notation $((n,K,d))$ to refer to such a code.
Two quantum codes are equivalent if they can be mapped into each other
by a permutation of the qubits combined with unitary transformations
confined to each qubit.

To verify that a code has minimum distance $d$, it suffices to restrict
one's attention to errors of the form
$$
\sigma_1\otimes \sigma_2\otimes \cdots \otimes \sigma_n,
$$
where each $\sigma_i$ ranges over the set
$$
\left\{ \pmatrix 1&0\\ 0&1\endpmatrix\!,
        \sigma_x=\pmatrix 0&1\\ 1&0\endpmatrix\!,
        \sigma_y=\pmatrix 0&-i\\ i&0\endpmatrix\!,
        \sigma_z=\pmatrix 1&0\\ 0&-1\endpmatrix \right\};
$$
we will denote the set of such errors by $\Cal{E}$.  (A proof of this fact
is given below using weight enumerators; see also [\knilllaflamme].)  For an
error $E$ in $\Cal{E}$, we define the weight $\wt(E)$ of $E$ as the number
of the $\sigma_i$ not equal to the identity.  Also, $\supp(E)$ is the
subset of $\{1,2,\ldots n\}$ consisting of the indices for which
$\sigma_i\ne 1$.  Thus $\wt(E)=|\supp(E)|$.

We will also need the following fact:

\proclaim{Lemma 1} Let $M$ be any operator on $V$.  Then $M$ can be
written as the following linear combination of the elements of $\Cal{E}$:
$$
M={1\over 2^n}\sum_{E\in \Cal{E}} \Tr(M E) E.
$$
\endproclaim

\demo{Proof}
Note, first, that if $E,E'\in\Cal{E}$, then
$$
\Tr(E E')=\cases 2^n&\text{$E=E'$} \\
                 0  &\text{otherwise} \endcases
$$
Thus $\{2^{-n/2} E\mid E\in \Cal{E}\}$ gives an orthonormal basis of
$\Hom(V,V)$, and the result follows immediately.
\qed\enddemo

The Shor-Laflamme enumerators are defined (up to a normalization factor
which we omit):
$$
\eqalign{
A_d(M_1,M_2)&=\sum_{E\in \Cal{E}\atop \wt(E)=d} \Tr(E M_1)\Tr(E M_2)\cr
B_d(M_1,M_2)&=\sum_{E\in \Cal{E}\atop \wt(E)=d} \Tr(E M_1 E M_2),\cr}
$$
where $M_1$ and $M_2$ are operators on $V$.

\proclaim{Theorem 2}
Let $P$ be the orthogonal projection onto a quantum code $\Cal{C}$ of
dimension $K$.  Then
$$
K B_i(P,P)\ge A_i(P,P)\ge 0
$$
for $0\le i\le n$.
\endproclaim

\demo{Proof}
(An alternate proof is given in [\shorlaflamme].)

First, note that
$$
A_d(M,M^\dagger)=\sum_{E\in \Cal{E}\atop \wt(E)=d} |\Tr(E M)|^2,
$$
so in particular, $A_d(P,P)\ge 0$.

Now, let $v$ be a random unit vector from $\Cal{C}$, uniformly distributed,
and consider
$$
E(A_d(v v^\dagger-{1\over K} P,v v^\dagger-{1\over K} P))\ge 0.
$$
In general, let $O$ be any operator on $V$, and consider
$$
\eqalign{
E(|\Tr(O v v^\dagger)-{1\over K} \Tr(O P)|^2)
&=
E(|\langle v| O| v\rangle|^2)-{1\over K^2}|\Tr(O P)|^2\cr
&=
{1\over K(K+1)}\left(|\Tr(OP)|^2+
                     \Tr(O P O^\dagger P)\right)\cr
&\phantom{={}}
-{1\over K^2} |\Tr(O P)|^2\cr
&=
{1\over K^2(K+1)} \left(K\Tr(O P O^\dagger P)-|\Tr(O P)|^2\right)\cr}
$$
Consequently,
$$
E(A_d(v v^\dagger-{1\over K} P,v v^\dagger-{1\over K} P))
=
{1\over K^2(K+1)} \left(K B_d(P,P)-A_d(P,P)\right),
$$
and
$$
K B_d(P,P)\ge A_d(P,P).
$$
\qed\enddemo

We will also have occasion to use enumerators $A_S(M_1,M_2)$ and
$B_S(M_1,M_2)$, where $S\subset\{1,2,\ldots n\}$:
$$
\eqalign{
A_S(M_1,M_2)&=\sum_{E\in \Cal{E}\atop \supp(E)=S} \Tr(M_1 E)\Tr(M_2 E)\cr
B_S(M_1,M_2)&=\sum_{E\in \Cal{E}\atop \supp(E)=S} \Tr(M_1 E M_2 E)\cr}
$$
Clearly theorem 2 applies to these enumerators as well.

Finally, we consider two polynomials
$$
\eqalign{
A(x,y)&=\sum_{0\le d\le n} A_d(M_1,M_2) x^{n-d}y^d\cr
B(x,y)&=\sum_{0\le d\le n} B_d(M_1,M_2) x^{n-d}y^d\cr}
$$

\head 2. The unitary enumerators ($A'$ and $B'$) \endhead

One problem with the Shor-Laflamme enumerators as defined is that,
while they are indeed invariants of the code under the $U(2)$ action
on each qubit ([\shorlaflamme], also see below), this is not immediately
obvious from their definition.  This motivates the introduction of
two new enumerators, $A'$ and $B'$:
$$
\eqalign{
A'_S(M_1,M_2)&=2^{|S|} E_{U_S} \Tr(M_1 U_S)\Tr(M_2 U_S^\dagger)\cr
B'_S(M_1,M_2)&=2^{|S|} E_{U_S} \Tr(M_1 U_S M_2 U_S^\dagger),\cr}
$$
where $U_S$ is a (uniformly) random unitary operator on the qubits
indexed by $S$.  These are clearly invariant under any equivalence
that maps qubits in $S$ to qubits in $S$.  Similarly, we define:
$$
\eqalign{
A'_d(M_1,M_2)&=\sum_{|S|=d} A'_S(M_1,M_2)\cr
B'_d(M_1,M_2)&=\sum_{|S|=d} B'_S(M_1,M_2).\cr}
$$
These are clearly invariants under equivalence.  We also consider
polynomials $A'(x,y)$ and $B'(x,y)$, defined in the obvious way.

The new enumerators have the following simpler definitions:

\proclaim{Theorem 3} Let $M_1$ and $M_2$ be any operators on $V$. Then
$$
\eqalign{
A'_S(M_1,M_2)&=\Tr_S(\Tr_{S^c}(M_1)\Tr_{S^c}(M_2))\cr
B'_S(M_1,M_2)&=\Tr_{S^c}(\Tr_S(M_1)\Tr_S(M_2)),\cr}
$$
where $S^c$ denotes the complement of $S$.  In particular,
$$
A'_S(M_1,M_2)=B'_{S^c}(M_1,M_2).
$$
\endproclaim

\demo{Proof}
We will use the following facts about random unitary matrices:
$$
\eqalign{
E_U(\Tr(A U)\Tr(B U^\dagger))&={1\over \dim(U)} \Tr(AB)\cr
E_U(U A U^\dagger)&={1\over \dim(U)} \Tr(A),\cr}
$$
both of which follow easily from the fact that $\Tr$ is an irreducible
character of the unitary group.

Now, then, we have:
$$
\eqalign{
A'_S(M_1,M_2)
&=2^{|S|} E_{U_S} \Tr(M_1 U_S)\Tr(M_2 U_S^\dagger)\cr
&=2^{|S|} E_{U_S} \Tr(Tr_{S^c}(M_1) U_S)\Tr(\Tr_{S^c}(M_2) U_S^\dagger)\cr
&=\Tr(Tr_{S^c}(M_1)\Tr_{S^c}(M_2)).\cr}
$$
For $B'_S$, the proof is slightly more complicated.  The crucial
observation is that
$$
\eqalign{
B'_S(M_1,M_2)&=B'_S(E_{U_S}(U_S M_1 U_S^\dagger),
                    E_{U_S}(U_S M_2 U_S^\dagger))\cr
&=2^{-2|S|} B'_S(\Tr_S(M_1)\otimes 1_S,
                 \Tr_S(M_2)\otimes 1_S)\cr
&=2^{-|S|} \Tr((\Tr_S(M_1)\otimes 1_S)
               (\Tr_S(M_2)\otimes 1_S))\cr
&=\Tr_{S^c}(\Tr_S(M_1)\Tr_S(M_2)).\cr}
$$
\qed\enddemo

These new enumerators are closely related to the Shor-Laflamme enumerators:

\proclaim{Theorem 4} Let $M_1$ and $M_2$ be any operators on 
$V$.  Then
$$
\eqalign{
A'_S(M_1,M_2)&=2^{-|S|} \sum_{T\subset S} A_T(M_1,M_2)\cr
B'_S(M_1,M_2)&=2^{-|S|} \sum_{T\subset S} B_T(M_1,M_2).\cr}
$$
\endproclaim

\demo{Proof}
Expand $U_S$ in terms of $\Cal{E}$:
$$
\eqalign{
A'_S(M_1,M_2)
&=
2^{|S|} E_{U_S} \Tr(M_1 U_S)\Tr(M_2 U_S^\dagger)\cr
&=
2^{|S|} 2^{-2n} \sum_{E_1,E_2\in \Cal{E}}
         E_{U_S} (\Tr(U_S E_1)\Tr(U_S^\dagger E_2))
         \Tr(M_1 E_1)\Tr(M_2 E_2)\cr}
$$
and similarly for $B'_S(M_1,M_2)$.  Thus, we need to compute
$$
2^{|S|} E_{U_S}(\Tr(U_S E_1)\Tr(U_S^\dagger E_2))=A_S(E_1,E_2).
$$
Now, by theorem 3,
$$
A'_S(E_1,E_2)
=\Tr_S(\Tr_{S^c}(E_1)\Tr_{S^c}(E_2))
$$
If $\supp(E_1)\not\subset S$, then $\Tr_{S^c}(E_1)=0$; otherwise,
if both $\supp(E_1)$ and $\supp(E_2)\subset S$, then $A'_S(E_1,E_2)=0$
unless $E_1=E_2$, when
$$
A'_S(E_1,E_2)=2^{|S|} 2^{2(n-|S|)}=2^{2n-|S|}
$$
Thus
$$
\eqalign{
A'_S(M_1,M_2)&=2^{-|S|} \sum_{E\in \Cal{E}\atop \supp(E)\subset S}
                             \Tr(M_1 E)\Tr(M_2 E)\cr
             &=2^{-|S|} \sum_{T\subset S} A_T(M_1,M_2),\cr}
$$
and similarly for $B'$.
\qed\enddemo

\proclaim{Corollary 5} Let $M_1$ and $M_2$ be operators on $V$.  Then
$$
\eqalign{
A'_d(M_1,M_2)&=2^{-d} \sum_{0\le i\le d} {n-i\choose n-d} A_i(M_1,M_2)\cr
B'_d(M_1,M_2)&=2^{-d} \sum_{0\le i\le d} {n-i\choose n-d} B_i(M_1,M_2).\cr}
$$
Similarly,
$$
\eqalign{
A'(x,y)&=A(x+y/2,y/2)\cr
B'(x,y)&=B(x+y/2,y/2).\cr
}
$$
\endproclaim

\demo{Proof}
We have
$$
\eqalign{
A'_d(M_1,M_2)&=\sum_{|S|=d} A'_S(M_1,M_2)\cr
             &=2^{-d} \sum_{|S|=d} \sum_{T\subset S} A_T(M_1,M_2)\cr
             &=2^{-d} \sum_{0\le i\le d} \sum_{|T|=i}
                                         \sum_{T\subset S\atop |S|=d}
                             A_T(M_1,M_2)\cr
             &=2^{-d} \sum_{0\le i\le d} \sum_{|T|=i}
                                  {n-i\choose n-d}
                             A_T(M_1,M_2)\cr
             &=2^{-d} \sum_{0\le i\le d} 
                                  {n-i\choose n-d}
                             A_i(M_1,M_2),\cr}
$$
and similarly for $B'$.

For the enumerator polynomials, we have:
$$
\eqalign{
A'(x,y)&=\sum_{0\le d\le n} A'_d x^{n-d}y^d\cr
       &=\sum_{0\le d\le n} 2^{-d} \sum_{0\le i\le d} 
                  {n-i\choose n-d} A_i x^{n-d} y^d\cr
       &=\sum_{0\le i\le n} A_i \sum_{i\le d\le n}
                         2^{-d} {n-i\choose n-d} x^{n-d} y^d\cr
       &=\sum_{0\le i\le n} A_i \sum_{i\le d\le n}
                         2^{-d} {n-i\choose n-d} x^{n-d} y^d\cr
       &=\sum_{0\le i\le n} A_i \left({y\over 2}\right)^i\left(x+{y\over
       2}\right)^{n-i},\cr}
$$
and similarly for $B'$.
\qed\enddemo

\proclaim{Corollary 7}
The enumerators $A$ and $B$ are invariants under equivalence.
\endproclaim

\demo{Proof}
The quantities $A_d$ and $B_d$ are fixed linear combinations of the 
manifestly invariant quantities $A'$ and $B'$.
\qed\enddemo

Recall from theorem 3 that $A'_S=B'_{S^c}$, and thus $A'_d=B'_{n-d}$ and
$A'(x,y)=B'(y,x)$.  This implies the following relationship between
$A(x,y)$ and $B(x,y)$:

\proclaim{Theorem 7. (Quantum MacWilliams identities)}
Let $A$ and $B$ be the Shor-Laflamme enumerators associated to a pair
of operators $M_1$ and $M_2$.  Then
$$
A(x,y)=B({x+3y\over 2},{x-y\over 2}).
$$
\endproclaim

\demo{Proof}
(see also [\shorlaflamme])
$$
A(x,y)=A'(x-y,2y)=B'(2y,x-y)=B(2y+{x-y\over 2},{x-y\over 2})
                            =B({x+3y\over 2},{x-y\over 2}).
$$
\qed\enddemo

\proclaim{Theorem 8}
Let $\Cal{C}$ be a quantum code of dimension $K$, with associated
projection $P$.  Then for $0\le i\le n$,
$$
K B'_i(P,P)\ge A'_i(P,P).
$$
If $K B'_{d-1}(P,P)=A'_{d-1}(P,P)$, then $\Cal{C}$ has minimum
distance at least $d$.  If $K=1$, then $B'_i(P,P)=A'_i(P,P)$ for all $i$.
\endproclaim

\demo{Proof}
By the same proof as for theorem 2, we have:
$$
K B'_i(P,P)-A'_i(P,P)=
K^2(K+1) E_{v\in \Cal{C}} A'_i(vv^\dagger-P,vv^\dagger-P).
$$
Consequently,
$$
K B'_i(P,P)-A'_i(P,P)\ge 0,
$$
with equality only when
$$
E_{v\in \Cal{C}} |\langle v|U_i|v\rangle-{1\over K}\Tr(U_i P)|^2=0,
$$
where $U_i$ ranges over all $i$-qubit errors.  But this
expectation is simply a variance; consequently, it is 0 precisely when
$$
\langle v|U_i|v\rangle=\langle w|U_i|w\rangle,
$$
for $v$ and $w$ ranging over all unit vectors in $\Cal{C}$.
This is precisely the condition that $\Cal{C}$ have minimum distance
$i+1$.

Finally, if $K=1$, then we have:
$$
B'_i(P,P)\ge A'_i(P,P)=B'_{n-i}(P,P)\ge A'_{n-i}(P,P)=B'_i(P,P),
$$
so $B'_i(P,P)=A'_i(P,P)$.
\qed\enddemo

\proclaim{Corollary 9}
Let $\Cal{C}$ be a quantum code of dimension $K$, with associated
projection $P$.  Then $\Cal{C}$ has minimum distance at least $d$ if and
only if
$$
KB_i(P,P)=A_i(P,P)
$$
for $0\le i<d$.
\endproclaim

\demo{Proof}
The quantity
$$
K B'_{d-1}(P,P)-A'_{d-1}(P,P)
$$
is a positive linear combination of
$$
K B_i(P,P)-A_i(P,P)
$$
for $0\le i<d$; the result follows immediately.
\qed\enddemo

We also have the following result:

\proclaim{Theorem 10}
Let $\Cal{C}$ be a quantum code of dimension $K$, with associated
projection $P$.  Then
$$
KB'_S(P,P)=A'_S(P,P)
$$
if and only if
$$
\Tr_{S^c}(v v^\dagger)
$$
is constant when $v$ ranges over unit vectors in $\Cal{C}$.
\endproclaim

\demo{Proof}
As before, we have
$$
K B'_S(P,P)-A'_S(P,P)\propto
E_{v\in \Cal{C}} A'_S(v v^\dagger-{1\over K} P,v v^\dagger-{1\over K} P)
$$
So equality holds if and only if
$$
E_{v\in \Cal{C}} |\Tr_{S^c}(v v^\dagger-{1\over K} P)|^2=0,
$$
or
$$
\Tr_{S^c}(v v^\dagger)={1\over K} \Tr_{S^c}(P)
$$
for all unit vectors $v\in\Cal{C}$.
\qed\enddemo

This result has the following physical interpretation: $K
B'_S(P,P)=A'_S(P,P)$ if and only if the code $\Cal{C}$ can correct for the
erasure of the qubits in $S$; the qubits in $S$ alone carry no information
about the encoded state.  (Such errors can occur, for instance, in
photon-based implementations of quantum computers, in which occasionally a
photon is lost.)  Consequently, we have the following result:

\proclaim{Theorem 11}
A quantum code $\Cal{C}$ has minimum distance $d$ if and only if it
can correct for any erasure of size $d-1$.
\endproclaim

Remark.  When talking about correcting for erasures, the assumption is that
it is known which qubits have been erased.  The point of this theorem is
that it is generally easier to give an algorithm for correcting erasures
than to give an algorithm for correcting ordinary errors; see, for
instance, theorem 21 below.

Remark.  This theorem is the analogue of a theorem for classical error
correcting codes ([\macwilliamssloane]).

\head 3. Enumerators for codes of block size greater than 2\endhead

We now wish to generalize everything to codes with block size greater than
2.  That is, we replace the state space $V$ by a tensor product of $n$
Hilbert spaces $V_1$ through $V_n$, with $\dim(V_i)=D_i$, not necessarily
equal to 2; in general, we will not even assume that the $V_i$ all have the
same dimension.  A quantum code is again a subspace $\Cal{C}$ of $V$.

Clearly, the unitary enumerators extend directly to this case:

\proclaim{Definition}  Let $S$ be any subset of $\{1,2,\ldots n\}$,
and let $M_1$ and $M_2$ be any operators on $V$.  Then define
$$
\eqalign{
A'_S(M_1,M_2)&=\Tr_S(\Tr_{S^c}(M_1)\Tr_{S^c}(M_2))\cr
B'_S(M_1,M_2)&=\Tr_{S^c}(\Tr_S(M_1)\Tr_S(M_2)).\cr}
$$
\endproclaim

We can also define these as we did for binary codes:

\proclaim{Theorem 12}
For any $S$ and operators $M_1$, $M_2$,
$$
\eqalign{
A'_S(M_1,M_2)&=\dim(V_S) E_{U_S} (\Tr(M_1 U_S)\Tr(M_2 U^\dagger_S)),\cr
B'_S(M_1,M_2)&=\dim(V_S) E_{U_S} (\Tr(M_1 U_S M_2 U^\dagger_S)).\cr}
$$
\endproclaim

\demo{Proof}
The proof of theorem 3 carries over directly.\qed\enddemo

To generalize $A_S$ and $B_S$, it will be convenient to introduce yet
another definition of $A'_S$ and $B'_S$.  For an operator $M$ on $V$,
define new operators $M'_S$ and $M_S$ for all $S\subset \{1,2,\ldots n\}$:
$$
\eqalign{
M'_S&={1\over \dim(V_{S^c})} (\Tr_{S^c}(M)\otimes 1_{S^c})\cr
M_S&=\sum_{T\subset S} (-1)^{|S|-|T|} M'_T.\cr}
$$

\proclaim{Theorem 12}
For all operators $M$, $N$,
$$
\eqalign{
A'_S(M,N)&=\dim(V_{S^c}) \Tr(M'_S N'_S)\cr
B'_S(M,N)&=\dim(V) E_{U\in U(V)} A'_S(M U,N U^\dagger)\cr}
$$
\endproclaim

\demo{Proof}
We have:
$$
\eqalign{
\Tr(M'_S N'_S)&=
{1\over \dim(V_{S^c})^2}
\Tr((\Tr_{S^c}(M)\otimes 1_{S^c})(\Tr_{S^c}(N)\otimes 1_{S^c}))\cr
&=
{1\over \dim(V_{S^c})}
\Tr(\Tr_{S^c}(M)\Tr_{S^c}(N)).\cr}
$$

The statement about $B'_S$ follows easily from theorem 11.
\qed
\enddemo

We can now define $A_S$ and $B_S$:

\proclaim{Definition}
Let $S$ be any subset of $\{1,2,\ldots n\}$,
and let $M$ and $M$ be any operators on $V$.  Then define
$$
\eqalign{
A_S(M,N)&=\dim(V) \Tr(M_S N_S),\cr
B_S(M,N)&=\dim(V) E_{U\in U(V)} A_S(M U,N U^\dagger).\cr}
$$
\endproclaim

To see how this relates to $A'$ and $B'$, we will need the following
results:

\proclaim{Lemma 13}
The map $M\mapsto M'_S$ is an orthogonal projection on $\Hom(V,V)$ for
all $S$.  Moreover,
$$
(M'_S)'_T=M'_{S\cap T}.
$$
\endproclaim

\demo{Proof}
Let us first show that
$$
(M'_S)'_T=M'_{S\cap T}.
$$
But
$$
\eqalign{
(M'_S)_T&={1\over \dim(V_{T^c})} (\Tr_{T^c}(M'_S)\otimes 1_{T^c})\cr
&={1\over \dim(V_{T^c})\dim(V_{S^c})}
(\Tr_{T^c}(\Tr_{S^c}(M)\otimes 1_{S^c})\otimes 1_{T^c})\cr
&={\dim(V_{S^c\cap T^c})\over \dim(V_{T^c})\dim(V_{S^c})}
\Tr_{S^c\cup T^c}(M)\otimes 1_{S^c\cup T^c}\cr
&=M'_{S\cap T}.\cr}
$$

In particular, $(M'_S)_S=M'_S$.
It remains only to show that if $M_S=M$ and $N_S=0$, then
$$
\Tr(MN)=0.
$$
But $N_S=0$ if and only if $\Tr_{S^c}(N)=0$.  Thus
$$
\eqalign{
\Tr(MN)
&=\Tr(M_S N)\cr
&=\Tr((\Tr_{S^c}(M)\otimes 1_{S^c}) N)\cr
&=\Tr(\Tr_{S^c}(M)\Tr_{S^c}(N))\cr
&=0.\cr}
$$
\qed\enddemo

\proclaim{Corollary 14}
For all $S\subset \{1,2,\ldots n\}$, the map $M\mapsto M_S$ is
an orthogonal projection.  Moreover,
$$
\Tr(M_S N_T)=0
$$
unless $S=T$, when
$$
\Tr(M_S N_S)= \sum_{R\subset S} \Tr(M'_R N'_R) (-1)^{|S|-|R|}.
$$
Finally,
$$
M'_S=\sum_{T\subset S} M_T.
$$
\endproclaim

\demo{Proof}
This follows readily from theorem 13 and the M\"obius inversion formula.
\qed\enddemo

\proclaim{Theorem 15}
Let $M$ and $N$ be any operators on $V$.  Then
$$
\eqalign{
A'_S(M,N)&={1\over \dim(V_S)} \sum_{T\subset S} A_T(M,N),\cr
B'_S(M,N)&={1\over \dim(V_S)} \sum_{T\subset S} B_T(M,N),\cr
A_S(M,N)&=\sum_{T\subset S} (-1)^{|S|-|T|} \dim(V_T) A'_T(M,N),\cr
B_S(M,N)&=\sum_{T\subset S} (-1)^{|S|-|T|} \dim(V_T) B'_T(M,N).\cr
}
$$
\endproclaim

\demo{Proof}
We have:
$$
\eqalign{
A'_S(M,N)&=\dim(V_{S^c}) \Tr(M'_S N'_S)\cr
         &=\dim(V_{S^c}) \sum_{T,T'\subset S} \Tr(M_T N_{T'})\cr
         &=\dim(V_{S^c}) \sum_{T\subset S} \Tr(M_T N_T)\cr
         &={1\over \dim(V_S)} \sum_{T\subset S} A_T(M,N).\cr}
$$
The remaining results follow similarly.
\qed\enddemo

\proclaim{Theorem 16}
Let $\Cal{C}$ be a quantum code of dimension $K$ in $V$, and let $P$ be its
associated projection.  Then for all $S\subset \{1,2,\ldots n\}$,
$$
\eqalign{
K B'_S(P,P)&\ge A'_S(P,P)\ge 0,\cr
K B_S(P,P)&\ge A_S(P,P)\ge 0.\cr}
$$
In particular, if $K=1$, then $A'_S(P,P)=B'_S(P,P)$ and $A_S(P,P)=B_S(P,P)$.
\endproclaim

\demo{Proof}
The proof for $A'$ and $B'$ proceeds as in theorem 8; it remains only
to consider $A$ and $B$.

First, let $M$ be any operator on $V$, and observe that
$$
A_S(M,M^\dagger)=\dim(V) \Tr(M_S M_S^\dagger)\ge 0.
$$

Now, let $v$ be a uniformly randomly chosen unit vector from $\Cal{C}$.
Then
$$
E_{v\in \Cal{C}} A_S(v v^\dagger-{1\over K} P,v v^\dagger-{1\over K} P)
=
{1\over K^2(K+1)} (K B_S(P,P)-A_S(P,P));
$$
this is just a linear combination of the corresponding equations for $A'$
and $B'$.  The theorem follows.
\qed\enddemo

Let us now assume that $D_i=D$ for all $i$.  Then it makes sense to
consider $A_d=\sum_{|S|=d} A_S$, and so on.

\proclaim{Theorem 17}
Let $\Cal{C}$ be a quantum code of dimension $K$ in $V$, with associated
projection $P$.  Then $\Cal{C}$ has minimum distance at least $d$ if
and only if
$$
K B_i(P,P)=A_i(P,P)
$$
for $0\le i<d$.
\endproclaim

\demo{Proof}
Clearly, $\Cal{C}$ has minimum distance at least $d$ if and only if
$$
K B'_{d-1}(P,P)=A'_{d-1}(P,P).
$$
But $K B'_{d-1}(P,P)-A'_{d-1}(P,P)$ is a positive linear combination of
$K B_i(P,P)-A_i(P,P)$ for $0\le i\le d-1$.
\qed\enddemo

We also have a MacWilliams transform:

\proclaim{Theorem 18}
Let $A$, $B$, $A'$, and $B'$ be the polynomial enumerators associated
with a quantum code $\Cal{C}$.  Then
$$
\eqalign{
A'(x,y)&=B'(y,x)\cr
A(x,y)&=B({x+(D^2-1)y\over D},{x-y\over D})\cr}
$$
\endproclaim

\demo{Proof}
The first assertion follows by inspection.  For the second assertion,
we note that
$$
\eqalign{
A'(x,y)&=A(x+{y\over D},{y\over D}),\cr
B'(x,y)&=B(x+{y\over D},{y\over D}),\cr
A(x,y)&=A'(x-y,Dy),\cr
B(x,y)&=B'(x-y,Dy),\cr}
$$
so
$$
\eqalign{
A(x,y)&=A'(x-y,Dy)=B'(Dy,x-y)\cr
  &=B(Dy+{x-y\over D},{x-y\over D})=B({x+(D^2-1)y\over D},{x-y\over D}).\cr}
$$
\qed\enddemo

\head 4. Shadow enumerators (conjecture) \endhead

For binary codes, there is an additional enumerator to consider,
namely the shadow enumerator ([\shadowenum]), which can be defined
by
$$
S(x,y)=A({x+3y\over 2},{x-y\over 2}).
$$
If one writes $S(x,y)$ in terms of $A'$, something rather curious happens:
$$
S(x,y)=A'(x+y,y-x).
$$
This suggests the following definition for arbitrary block sizes:

\proclaim{Definition} Let $T$ be any subset of $\{1,2,\ldots n\}$,
and let $M$, $N$ be operators on $V$.  Then the {\it shadow enumerator}
of $M$ and $N$ is defined by
$$
S_T(M,N)=\sum_{R\subset \{1,2,\ldots n\}} (-1)^{|R\cap T^c|} A'_R(M,N).
$$
\endproclaim

The conjecture is then that $S_T(M,N)\ge 0$ whenever $M$ and $N$ are
positive semi-definite Hermitian operators (the case when $D_i=2$ for all
$i$ was essentially proved in [\shadowenum]). More explicitly:

\proclaim{Conjecture} Let $V=V_1\otimes V_2\otimes\cdots\otimes V_n$,
where $V_1$ through $V_n$ are Hilbert spaces.  Let $T$ be any 
subset of $\{1,2,\ldots n\}$, and let $M$ and $N$ be positive
semi-definite Hermitian operators on $V$.  Then
$$
\sum_{S\subset \{1,2,\ldots n\}} (-1)^{|S\cap T|}
\Tr(\Tr_{S^c}(M)\Tr_{S^c}(N))
\ge 0.
$$
\endproclaim

When $n=1$, this becomes
$$
|\Tr(M)\Tr(N)|\ge|\Tr(MN)|,
$$
which is easy to verify.

\head 5. Constructions for quantum codes \endhead

We will now use the unitary enumerators to examine some constructions of
new quantum codes from existing quantum codes.  We will assume that $D_i$
is constant ($=D$) throughout.

\proclaim{Theorem 19}
Suppose $\Cal{C}$ is a pure $((n,K,d))$ with $n,d\ge 2$.  Then there exists
a pure $((n-1,DK,d-1))$.
\endproclaim

\demo{Proof}
Let $P$ be the projection associated with $\Cal{C}$, and let
$P'=D\Tr_{\{1\}}(P)$.  The claim is then that $P'$ is the projection
associated with the desired code.

First, note that
$$
\Tr({P'}^2)=D^2 B'_{\{1\}}(P)=DK=\Tr(P').
$$
This, combined with the fact that $P'$ has at most $DK$ distinct
eigenvalues, implies that $P'$ is a projection.

It remains to show that $P'$ is pure of minimum distance $d-1$.  Thus,
let $S$ be a set of size $d-2$ in $\{2,3,\ldots n\}$, and observe:
$$
\eqalign{
B'_S(P',P')&=\Tr(\Tr_S(P')^2)\cr
          &=D^2\Tr(\Tr_{S\cup \{1\}}(P)^2)\cr
          &=D^2 B'_{S\cup \{1\}}(P)\cr
          &=D^2 D^{1-d} K\cr
          &=D^{2-d} (DK).\cr}
$$
\qed\enddemo

If $K=1$, this construction is reversible:

\proclaim{Theorem 20}
Suppose $\Cal{C}$ is a quantum code with projection matrix $P$ of rank
$D$. Then there exists a code $\Cal{C}'$ with $P=D\Tr_{\{1\}}(P')$;
any two such codes are equivalent.  The new code has unitary enumerator
$$
A'_i(\Cal{C}')=D^{-2} (A'_i(\Cal{C})+B'_{i-1}(\Cal{C})).
$$
\endproclaim

\demo{Proof}
$P'$ clearly must have rank 1; consequently, we need a vector $v'$
with $P'=v' {v'}^\dagger$.

Since $\Tr_{\{1\}}(v' {v'}^\dagger)$ is a projection, it follows that
$v'$ must be writable in the form
$$
\sum_{0\le i<D} {1\over \sqrt{D}} w_i\otimes v_i,
$$
where $v_i$ ranges over some orthonormal basis for $\Cal{C}$,
and $w_i$ ranges over some orthonormal basis for $V_1$.
Conversely, any such $v'$ gives a suitable $P'$.  Uniqueness follows
from the fact that the freedom in the $w_i$ can be absorbed into the
freedom in the $v_i$, which in turn can be absorbed by applying an
element of $U(V_1)$.

Finally, let $S\subset\{1,2,\ldots,n\}$.  If $S$ does not contain
1, then:
$$
\eqalign{
A'_S(P',P')
&=
\Tr_S(\Tr_{S^c}(P')^2)\cr
&=
D^{-2} \Tr_S(\Tr_{S^c}(P)^2)\cr
&=
D^{-2} A_S(P,P)\cr}
$$
Now, if $S$ does contain 1, we have:
$$
\eqalign{
A'_S(P',P')&=B'_{S^c}(P',P')\cr
           &=A'_{S^c}(P',P')\cr
           &=D^{-2} A'_{S^c}(P,P)\cr
           &=D^{-2} B'_{S-\{1\}}(P,P).\cr}
$$
where the second and fourth equalities follow from theorem 16 and the fact
that $P'$ has rank 1.

The desired result follows by summing over $S$ of size $i$.
\qed\enddemo

Finally, let us consider concatenated codes.  Let $\Cal{C}_1$ be
a $((n_1,K_1,d_1))$, on blocks of size $D_1$, and let $\Cal{C}_2$ be
a $((n_2,D_1,d_2))$, on blocks of size $D_2$.  Then one can construct
a new code $\Cal{C}_2(\Cal{C}_1)$, by encoding each block of $\Cal{C}_1$
using $\Cal{C}_2$.  (Strictly speaking, the concatenated code also
depends on the specific encoding map used for $\Cal{C}_2$.)  Clearly,
the concatenated code encodes $K_1$ states in $n_1n_2$ blocks of size
$D_1D_2$; it remains only to consider its minimum distance:

\proclaim{Theorem 21}
Let $\Cal{C}_1$ and $\Cal{C}_2$ be as above.  Let
$\Cal{C}=\Cal{C}_2(\Cal{C}_1)$ be any concatenation of $\Cal{C}_1$ and
$\Cal{C}_2$.  Then $\Cal{C}$ has minimum distance at least $d_1d_2$.
\endproclaim

\demo{Proof}
By theorem 11 (which clearly holds for block codes as well), it suffices to
give an algorithm for correcting erasures of size $d_1d_2-1$.  Suppose,
therefore, that $d_1d_2-1$ blocks of $\Cal{C}$ have been erased.  The
correction algorithm is quite simple: decode the outer encoding, then
decode the inner encoding.

We can decode erasures of up to size $d_2-1$ in $\Cal{C}_2$.  Thus the only
blocks of the inner encoding that will be unrecoverable are those that
suffered at least $d_2$ erasures.  Clearly, there can be at most $d_1-1$
such blocks.  But this can be corrected, using the decoding algorithm for
$\Cal{C}_1$.
\qed\enddemo

\head Conclusion \endhead

We have furthered the enumerator theory of Shor and Laflamme, with the help
of two new manifestly invariant enumerators.  Since the definition of these
enumerators did not depend on the codes being binary, we could readily
extend the theory to quantum codes on larger alphabets.  We also used the
new enumerators to clarify the nature of the relationship between the
Shor-Laflamme enumerators, and to give a simpler condition for a quantum
code to have specified minimum distance.

\head Acknowledgments\endhead

We would like to thank P. Shor and N. Sloane for many helpful discussions;
we would also like to thank C. Bennett for a helpful discussion on
erasures.

\head References \endhead
\item{[\knilllaflamme]}
E. Knill and R. Laflamme, ``A theory of quantum error correcting codes'',
LANL e-print quant-ph/9604034.

\item{[\macwilliamssloane]}
F. J. MacWilliams and N. J. A. Sloane, The Theory of Error-Correcting
Codes, North-Holland, New York, 1977.

\item{[\shadowenum]}
E. M. Rains, ``Quantum shadow enumerators'', LANL e-print quant-ph/9611001.

\item{[\shorlaflamme]}
P.W. Shor and R. Laflamme, ``Quantum MacWilliams identities'',
LANL e-print quant-ph/9610040.
\enddocument
\bye